\documentclass[aps,prl,twocolumn,superscriptaddress]{revtex4}
\usepackage{makeidx}
\usepackage{graphicx}
\usepackage{amsmath}
\usepackage{amsfonts}
\usepackage{amssymb}
\usepackage{bm}
\usepackage{color}

\begin{document}

\title{Magnetization Currents of Fluctuative Cooper Pairs}
\author{A.V. Kavokin}
\affiliation{CNR-SPIN, Viale del Politecnico 1, I-00133 Rome, Italy}
\affiliation{Russian Quantum Center, 100, Novaya Str., Skolkovo, Moscow Region, 143025,
Russia}
\author{A.A.~Varlamov}
\affiliation{CNR-SPIN, Viale del Politecnico 1, I-00133 Rome, Italy}
\affiliation{Russian Quantum Center, 100, Novaya Str., Skolkovo, Moscow Region, 143025,
Russia}
\date{\today }

\begin{abstract}
Recent experiments show that the Nernst-Ettingshausen effect is orders of
magnitude stronger than the thermoelectric Seebeck effect in superconductors
above the critical temperature. We explain different magnitudes of the two
effects accounting for the magnetization current of virtual Cooper pairs.
The method allows for detailed understanding of the surprising non-monotonic
dependence of the Nernst-Ettingshausen coefficient on the magnetic field.
\end{abstract}

\maketitle

\emph{Introduction.} Thermoelectric and thermomagnetic phenomena in solids
were discovered in the XIXth century \cite{SXIX,NXIX}.The most significant
among them are the Seebeck effect (SE), the Nernst-Ettingshausen effect
(NEE) and the Ettingshausen effect (EE). The SE, also referred to as the
differential thermopower, consists in the induction of the electric field in
a conducting sample subjected to the gradient of temperature at zero
electric current (open circuit) condition. The field is induced in the
temperature gradient direction. The NEE consists in the induction of the
electric field $\mathbf{E}$\ in conducting samples subjected to a magnetic
field $\mathbf{H}$ and the temperature gradient $\mathbf{\nabla }T$ applied
in the perpendicular to $\mathbf{H}$\ direction. The electric field is
measured in the direction perpendicular to both magnetic field and
temperature gradient in the open circuit regime and adiabatic conditions
(both electric currents in the sample and the thermal flaw in the direction
of $\mathbf{E}$ equal to zero, $j_{x}=j_{y}=0$ and $q_{y}=0$, respectively).
In practice, the adiabatic condition is usually substituted by the isotermic
one: $\nabla _{y}T=0$\cite{K41}. The EE is reciprocal to NEE: it consists in
the induction of the temperature gradient $\nabla _{x}T$ if the current $%
j_{y}\neq 0$ propagates through the sample perpendicularly to the applied
magnetic field $H_{z}$ in the adiabatic conditions: $\nabla _{y}T=0$ and $%
j_{x}=q_{x}=0.$ Due to the Onsager principle of the transport coefficients
symmetry, NEE and EE usually are correlated.

The aforementioned phenomena found their theoretical explanation only in the
middle of the XXth century in the works by Mott \cite{M36} and Sondheimer 
\cite{S48}. In a degenerate Fermi gas, thermoelectric and thermomagnetic
effects were shown to be controlled by the particle-hole asymmetry, with the
magnitudes of SE, NEE and EE being governed by the factor $\sim T/E_{F},$
where $E_{F}$ is the Fermi energy. As a result, the Seebeck and
Nernst-Ettingshausen coefficients for good conductors at room temperatures
are of the order of $S=E_{x}/\nabla _{x}T\sim $\ $10^{-2}\div 10^{-1}$\ $\mu
V/K$ and $\nu =E_{x}/\left( H_{z}\cdot \nabla _{x}T\right) \sim
10^{-3}-10^{-2}\mu V/\left( K\cdot T\right) $, respectively, while they are
much larger in the case of half-metals and degenerated semiconductors.

The SE in type I superconductors occurs due to the transformation of normal
excitations into Cooper pairs at the edges of samples subjected to a
temperature gradient \cite{GGK73}. The EE which requires the magnetic field
penetration in a sample can be observed only in type II superconductors and
is due to the entropy transport governed by the motion of vortices. Still,
the magnitude of this effect remains as small as $10^{-4}$\ $\mu V/\left(
K\cdot T\right) $, which is why\ the studies of SE and EE in superconductors
had only the fundamental interest, initially.

Nowadays, the control of heat fluxes and minimization of related losses are
crucially important in nanoelectronics. This is why the thermoelectric and
thermomagnetic phenomena in nanostructures and new materials attracted much
attention in the recent years. First indications of a sizeable EE in a wide
range of temperatures in superconductors above the critical temperature were
reported by Palstra \textit{et} \textit{al} \cite{P90} (see also \cite%
{US90,KR91}) who detected it in the optimally doped YBCO samples at
temperatures up to 10 K above the phase transition. The discovery of a giant
EE (hundred times larger than its value in conventional metals) in the
pseudogap state of $La_{2-x}Sr_{x}CuO_{4}$ \cite{Ong00} was a next milestone
followed by the similar finding (with a 10$^{3}$ enhancement in magnitude in
the wide range of temperatures) in the low-temperature superconductor $%
Nb_{0,15}Si0_{0,85}$ \cite{Aub06}. These observations were especially
surprising in view of the previously recorded data on the magnitude of the
Seebeck coefficient in the fluctuative regime of superconductors, undergoing
a weak singular decrease close to $T_{\mathrm{c}}$ but remaining of the same
order of magnitude as in the normal phase above $T_{\mathrm{c}}$ \cite%
{H90,LRGC93,Ri94}.

The theoretical description of fluctuation contributions to the
thermoelectric and thermomagnetic coefficients remains complex and
controversial. Initially, the fluctuation contribution to the Seebeck
coefficient in 3D superconductor was studied by Maki \cite{M73} in the
framework of the time dependent Ginzburg-Landau equation, and it was found
to be negligibly small. After the discovery of the anomaly in the Seebeck
coefficient behavior close to $T_{\mathrm{c}}$ in monocrystals of $%
YBa_{2}Cu_{3}O_{7-\delta }$ \cite{H90} the problem was revisited both
phenomenologically \cite{UD91} and microscopically \cite{RS94}. Both papers
concluded that the fluctuation correction to the Seebeck coefficient\ $S_{%
\mathrm{fl}}$ is proportional to the degree of particle-hole asymmetry. It
logarithmically depends on temperature above $T_{\mathrm{c}}$ in the 2D
case: $S_{\mathrm{fl}}^{\left( 2\right) }\sim \left( T/E_{F}\right) \ln %
\left[ T_{\mathrm{c}}/\left( T-T_{\mathrm{c}}\right) \right] .$ In what
concerns the fluctuative EE at weak magnetic fields, it was initially
studied in the framework of the GL approach in the same Ref. \cite{UD91}. It
was shown that the Cooper pairs contribution to the Ettingshausen
coefficient does not contain the smallness induced by particle-hole
asymmetry and exhibits much stronger temperature dependence than the normal
phase contribution$:$ $\beta _{yx}^{\mathrm{fl}}\sim T_{\mathrm{c}}/\left(
T-T_{\mathrm{c}}\right) .$ After the new experimental findings of Ref. \cite%
{Ong00}, the problem was revisited in Ref. \cite{U02}, where the linear
response theory result of Ref. \cite{UD91} was reproduced and the importance
of the magnetization currents was emphasized.

\begin{figure}[tbp]
\includegraphics[width=1.0\columnwidth]{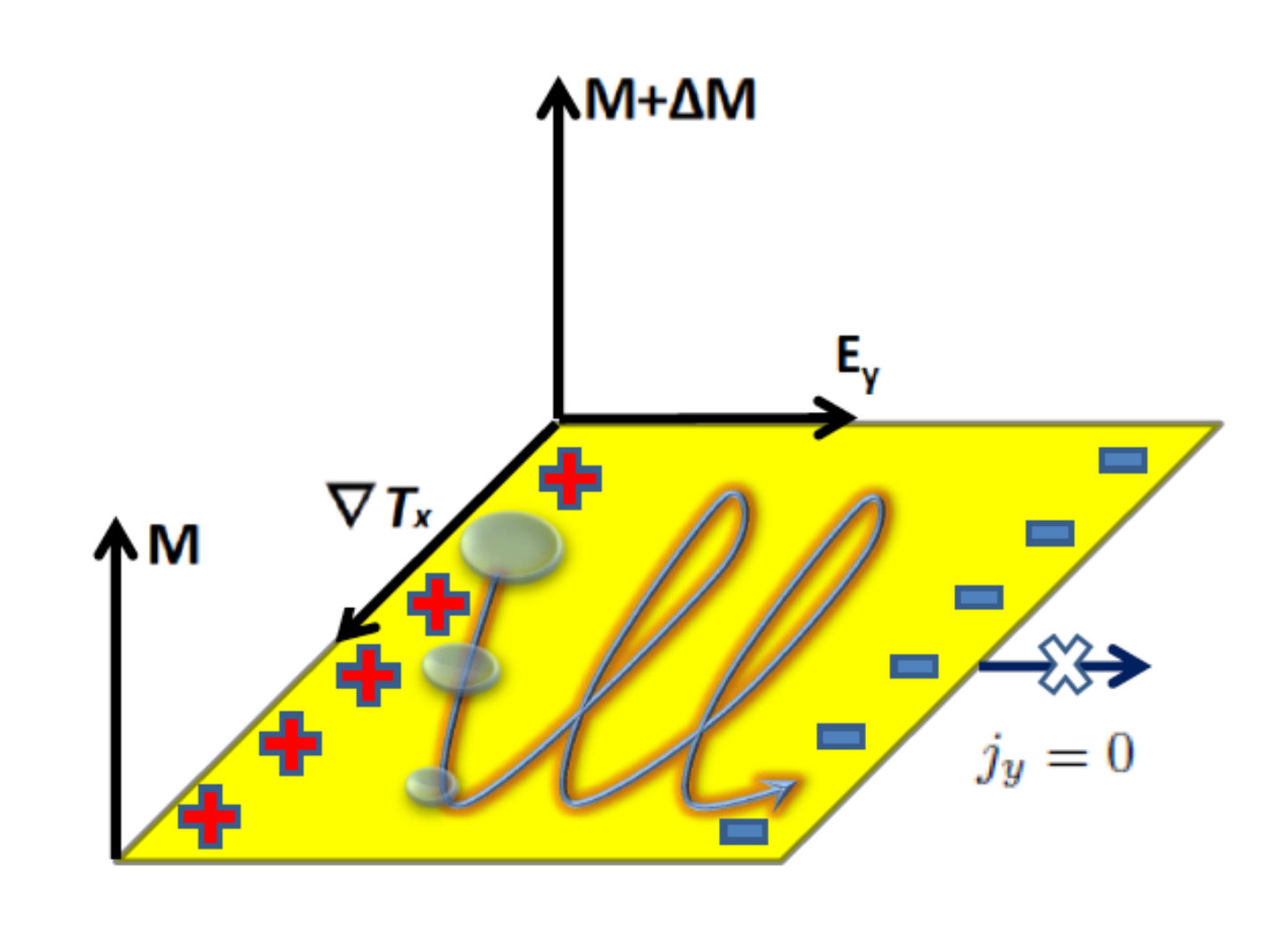}
\caption{Schematic representation of the FCP motion in the superconducting
film subjected to the temperature gradient along x-axis. The concentration
and size of FCP varies with temperature variation. The local magnetization
parallel to the external magnetic field varies along x-axis as well. The
spatial inhomogeneity of magnetization leads to the transformation of the
FCP trajectories from circular to trochoidal ones which is why the
magnetisation currents appear. To compensate these currents a voltage is
induced in y-direction that provides the main contribution to NEE.}
\label{magnetisation}
\end{figure}

The magnetization currents of virtual Cooper pairs are in the focus of the
Letter aimed at understanding of the surprisingly large difference in
magnitudes of SE\ and NEE in fluctuative superconductors. These currents may
be induced if the magnetization in the sample is spatially inhomogeneous.
Its inhomogeneity is caused by the temperature gradient. The induced
electric current contributes to the Ettingshausen coefficient (see the
schematic in Fig. \ref{magnetisation}). It can be easily expressed from the
Ampere law as $\mathbf{j}^{\mathrm{mag}}=\frac{c}{4\pi }\mathbf{\nabla }%
\times \mathbf{B},$ where $\mathbf{B}=\mathbf{H}+4\pi \mathbf{M}$, $\mathbf{H%
}$ is the spatially homogeneous external magnetic field, $\mathbf{M}$ is the
magnetization. In the presence of a temperature gradient $\nabla _{x}T$ one
can express the magnetization current as $j_{y}^{\mathrm{mag}}=$ -$c\left(
dM_{z}/dT\right) \nabla _{x}T$ \cite{YNO64,U02}. In the case of NEE, the
open circuit condition holds: $j_{x}=j_{y}=0$. In full analogy with a
classical Hall effect, the magnetization current in $y$ direction is
compensated by the induced Nernst-Ettingshausen voltage, which yields the
electric field $E_{y}^{mag}=\rho _{yy}j_{y}^{\mathrm{mag}}$ ($\rho _{yy}$ is
the diagonal component of the resistivity tensor, $\rho _{yy}=\rho _{xx}$).

The physics of magnetization currents has been revealed half a century ago
by Obraztsov \cite{YNO64} who noticed that the Onsager principle applied to
the thermoelectric tensor in the presence of magnetic fields can be
fulfilled only if these currents are accounted for. In normal metals the
magnetization currents are negligible so that they do not affect the
classical Sondheimer results \cite{S48}, obtained using the transport
equation approach.

In what concerns the properties of superconductors in the fluctuation
regime, Ussishikin et al. \cite{U02} demonstrated that accounting for the
contribution of magnetization currents to the heat flow in the vicinity of\ $%
T_{\mathrm{c}}$ one obtains the thrice lower value of the Ettingshausen
coefficient compared to what was predicted by Ullah and Dorsey \cite{UD91}.
The role of magnetization currents is even more important in the regime of
quantum fluctuations: the Kubo-like response contribution to the heat flow 
\cite{mah81} results in the violation of the third law of thermodynamics
which can only be rectified by taking into account the fluctuative Meissner
magnetization above $H_{\mathrm{c2}}\left( 0\right) $ \cite{SSVG09}. Similar
contradictions to the laws of thermodynamics were found in studies of
thermo-magnetic effects in other solid state systems \cite{HR87,GSV14}, and
in each case the magnetization currents contribution was crucial for
resolving the paradoxes. 

We present here a unified thermodynamic approach to NEE and SE in
fluctuative superconductor which elucidates the physics behind the striking
difference in their magnitudes. We show that while the Seebeck thermopower
is governed by the chemical potential temperature derivative, the NE effect
is dominated by magnetization currents of virtual Cooper pairs, which are
present in superconductors above the critical temperature \cite{LV09}. We
emphasize that we consider NEE, not EE, as the open circuit condition is
essential in our approach.

\emph{Generalities. }Let us consider a conductor subjected to a temperature
gradient and satisfying the boundary conditions: $j_{x}=j_{y}=0,$ $\nabla
_{y}T=0,$ and $\nabla _{x}T\neq 0$ (see Fig. 1). As a whole, it cannot be
characterized by a Fermi-Dirac distribution function as the unique
temperature cannot be defined throughout the volume of the sample. On the
other hand, the equilibrium distribution can be used for small enough
volumes where the temperature can be assumed constant. We shall work within
the\textit{\ local equilibrium }approximation introducing the locally
defined Fermi-Dirac function

\begin{equation}
f_{FD}\left( \varepsilon ,x\right) =\left\{ 1+\exp \left[ \left( \varepsilon
-\mu \left( x\right) \right) /k_{B}T\left( x\right) \right] \right\} ^{-1},
\label{local}
\end{equation}%
where $\mu \left( x\right) $ and $T\left( x\right) $ are the coordinate
dependent chemical potential and temperature. We underline that this
approximation is not universal. In particular, it is likely to fail if the
electron scattering is specifically energy dependent (Kondo effect,
thermoelectric effects in the vicinity of the $2\frac{1}{2}$ phase
transition) or in the case of a strong phonon drag effect. Nevertheless, it
remains a valid and powerful tool in a large variety of systems including
the most part of \textit{up-critical} (above the critical temperature $T_{c}$%
) superconductors. The conclusions of this Letter are restricted to the
systems where the aformentioned assumptions are applicable.

Once Eq. (\ref{local}) is valid, in the absence of electric current, the
electro-chemical potential $\mathcal{E}\left( x\right) $ must be constant
across the sample \cite{M96,A88}:

\begin{equation}
\mathcal{E}\left( x\right) =\mu \left( x\right) +e\varphi \left( x\right)
=const.  \label{electrochem}
\end{equation}%
It is instructive to apply to both parts of Eq. (\ref{electrochem}) the
gradient operator, having in mind that $E_{x}=-\triangledown \varphi .$ This
allows obtaining an important link between the temperature gradient and the
induced electric field: $eE_{x}=\left( \frac{d\mu }{dT}\right) \nabla T.$

Consequently, the Seebeck coefficient writes:

\begin{equation}
S\equiv \frac{E_{x}}{\nabla T}=\frac{1}{e}\frac{d\mu }{dT}.  \label{seemu}
\end{equation}

We will consider up-critical 2D suprconductors. The coexisiting subsystems
of fluctuation Cooper pairs (FCP) and electrons will be assumed locally
non-interacting. This means that in each small volume characterised by the
coordinate $x$\textbf{\ }the number of electrons involved in fluctuation
Cooper pairing is dependent only on $T\left( x\right) $. The FCP gas with
concentration $\mathcal{N}_{\mathrm{cp}}\left[ T\left( x\right) \right] $
and the depleted electron gas with the local concentration $n_{\mathrm{e}%
}\left( x\right) -\mathcal{N}_{\mathrm{cp}}\left[ T\left( x\right) \right]
/2 $ can be considered as two independent parallel conducting channels (the
indices $\mathrm{e}$ and $\mathrm{cp}$ are related to electrons and FCP,
respectively). Indeed, in the first order, the electron-electron
interactions in the Cooper channel are taken into account once the FCP
subsystem is introduced. In the second order, one would need to consider FCP
interactions, but these are known to be important only in the critical
vicinity of $T_{c}$.

The constancy of electrochemical potential condition applied to both
electron and Cooper pair subsystems results in:

\begin{equation}
E_{x}=-\triangledown \varphi _{\mathrm{e}}=-\triangledown \varphi _{\mathrm{%
cp}}=\frac{1}{2e}\triangledown \mu _{\mathrm{cp}}.
\end{equation}%
The chemical potential of FCP needs to be derived. From its general
definition one can write

\begin{equation}
\mu _{\mathrm{cp}}=\frac{\partial \mathcal{F}_{\mathrm{cp}}}{\partial 
\mathcal{N}_{\mathrm{cp}}}=\frac{\partial \mathcal{F}_{\mathrm{cp}}/\partial
\epsilon }{\partial \mathcal{N}_{\mathrm{cp}}/\partial \epsilon },
\label{def}
\end{equation}%
with $\mathcal{F}_{\mathrm{cp}}$ being the free energy of the FCP gas, $%
\epsilon =\ln \frac{T}{T_{c}}\approx \left( T-T_{c}\right) /T_{c}$ \ being
the reduced temperature. We shall specifically consider a superconducting
film of the thickness less than corresponding coherence length $\xi $.
Corresponding values of $\mathcal{F}_{\mathrm{cp}}^{\left( 2D\right) }$ and $%
\mathcal{N}_{\mathrm{cp}}^{\left( 2D\right) }$ \ above the critical
temperature can be written \ in the GL approximation \ as \cite{LV09} 
\begin{equation}
\mathcal{F}_{\mathrm{cp}}^{\left( 2D\right) }=-\frac{TS}{4\pi \xi ^{2}}%
\epsilon \ln \epsilon ;\;\mathcal{N}_{\mathrm{cp}}^{\left( 2D\right) }=\frac{%
1}{4\pi \alpha \xi ^{2}}\ln \frac{1}{\epsilon },  \label{2Ddef}
\end{equation}%
where the GL parameter $\alpha =4\pi ^{2}/\left[ 7\zeta \left( 3\right) %
\right] T_{c}/E_{F}$ is proportional to the electron-hole assymetry factor
(see Appendix A in Ref. \cite{LV09}). Substitution of Eq. (\ref{2Ddef}) to
Eq. (\ref{def}) results in \cite{Henni} 
\begin{equation}
\mu _{cp}=\alpha T\epsilon \ln \epsilon .  \label{mucp}
\end{equation}

\emph{Seebeck coefficient}. The substitution of Eq. (\ref{mucp}) in Eq. (\ref%
{seemu}) yields the Cooper pairs contribution to the Seebeck coefficient
above the superconducting transition:%
\begin{equation}
S_{\mathrm{cp}}=\frac{1}{2e}\frac{d\mu _{\mathrm{cp}}}{dT}=-\frac{1}{e}\frac{%
2\pi ^{2}T_{c}}{7\zeta \left( 3\right) E_{F}}\ln \frac{1}{\epsilon },
\label{See}
\end{equation}%
which exceeds the normal carrier contribution 
\begin{equation*}
S_{e}=\frac{1}{e}\frac{\partial \mu _{e}}{\partial T}=-\frac{\pi ^{2}T}{%
3eE_{F}},
\end{equation*}%
by a large logarithmic factor. We note that Eq. (\ref{See}) has a similar
temperature dependence to the previously published expressions (see Ref. 
\cite{M73,UD91,RS94,U02}) but, in contrast to them, the thermodynamic
approach provides us with the explicite value of the prefactor.

\emph{Nernst-Ettingshausen (NE) signal.} In a similar way, one can calculate
the contribution of FCP to the NE coefficient. It is dependent on the
components of the thermoelectric ($\beta _{ij})$ and resistivity ($\rho
_{ij} $) tensors as

\begin{equation}
\nu =\frac{\rho _{xx}\beta _{xy}+\rho _{xy}\beta _{yy}}{B}.  \label{nernst}
\end{equation}

Taking advantage of the relations $\beta _{yy}=-\frac{\sigma _{xx}}{2e}\frac{%
d\mu }{dT}$; $\beta _{xy}=c\frac{\partial M_{z}}{\partial T}$ derived in our
previous work \cite{VK2013} one finds from Eq. (\ref{nernst}) the NE\
coefficent of FCP as 
\begin{equation}
\nu _{\mathrm{cp}}=\frac{1}{2eH}\left( \frac{\sigma _{\mathrm{cp}}^{yx}}{%
\sigma _{\mathrm{cp}}^{xx}}\right) \frac{d\mu _{\mathrm{cp}}}{dT}+\frac{%
c\rho _{\mathrm{cp}}^{xx}}{H}\frac{dM_{\mathrm{cp}}^{z}}{dT}=\nu _{\mathrm{cp%
}}^{\left( \mathrm{th}\right) }+\nu _{\mathrm{cp}}^{\left( \mathrm{magn}%
\right) }.  \label{ncpfull}
\end{equation}%
The fluctuation contribution to the Hall conductivity $\sigma _{\mathrm{cp}%
}^{yx}$ is proportional to the coefficient of the electron-hole assymetry $%
T/E_{F}$ \ (see Ref. \cite{LV09}), which is why $\nu _{\mathrm{cp}}^{\left( 
\mathrm{th}\right) }$ appears to be small by a parameter $\alpha ^{2}$. On
the other hand, the magnetization term $\nu _{\mathrm{cp}}^{\left( \mathrm{%
magn}\right) }$ having the same singularity with respect to $\epsilon $ as $%
\nu _{\mathrm{cp}}^{\left( \mathrm{th}\right) },$ does not contain the small
factor $\alpha ^{2}$. This brings us to conclusion that the fluctuation
contribution to the NE coefficient is governed by the magnetization currents
of FCP. The magnetization term in (\ref{ncpfull}) is dependent on the
resistivity of FCP, who deviates from the result of Ussishkin \textit{et al }%
(Eq.(13) in Ref. \cite{U02}) who assumed that the NE\ coefficient is
dependent on the sum of normal phase and FCP conductivities in order to
achieve a good fit to the experimental data. We argue that considering
normal electrons and FCP as two parallel conductivity channels one may
derive the total voltage drop as product of current and resistivity in
either electronic or FCP channel. Therefore, the magnetization current of
FCP\ must be multiplied by the resistivity of FCP only.

Close to $T_{c}$\ and for weak enough magnetic fields, one can use the GL
approach \cite{LV09}. In the most interesting $2D$ case, the fluctuation
magnetization per unit area can be written as

\begin{equation*}
M_{\mathrm{cp}}^{\left( 2D\right) }(\epsilon ,h)=\frac{|e|T}{\pi }\left\{
\ln \frac{\Gamma (\frac{1}{2}+\frac{\epsilon }{2h})}{\sqrt{2\pi }}-\frac{%
\epsilon }{2h}\left[ \psi (\frac{1}{2}+\frac{\epsilon }{2h})-1\right]
\right\} ,
\end{equation*}%
while the longitudinal magnetoresistivity of FCP is given by the expression:%
\begin{equation*}
\rho _{\mathrm{cp}}^{\left( 2D\right) }\left( \epsilon ,h\right) =\frac{%
8h^{2}}{e^{2}\epsilon }\left[ \psi (\frac{1}{2}+\frac{\epsilon }{2h})-\psi (%
\frac{\epsilon }{2h})-\frac{h}{\epsilon }\right] ^{-1}.
\end{equation*}%
In the above expressions, $\psi (z)$ is the logarithmic derivative of the
Euler Gamma function $\Gamma (z)$, $h=2\pi \xi ^{2}H/\Phi _{0}$ is the
dimensionless magnetic field. Substituting these expressions into Eq. (\ref%
{ncpfull}) one finds for the NE signal $\left( N=\nu H\right) :$ 
\begin{eqnarray}
N_{\mathrm{cp}}^{\left( 2D\right) }\left( \epsilon ,h\right) &=&\left( \frac{%
4}{|e|}\right) \frac{\frac{\epsilon }{2h}\psi ^{\prime }\left( \frac{1}{2}+%
\frac{\epsilon }{2h}\right) -1}{1-\frac{\epsilon }{h}\left[ \psi (\frac{1}{2}%
+\frac{\epsilon }{2h})-\psi (\frac{\epsilon }{2h})\right] }  \notag \\
&=&\frac{8}{3|e|}\left( \frac{h}{\epsilon }\right) \left\{ 
\begin{array}{cc}
1, & h\ll \epsilon \\ 
\frac{3\epsilon }{2h}, & h\gg \epsilon%
\end{array}%
\right. .  \label{Nmagfull1}
\end{eqnarray}%
Eq. (\ref{Nmagfull1}) reproduces both the giant value of the fluctuation NE
signal observed in numerous experiments (as compared to $N_{\mathrm{e}}=-\pi
^{2}T\tau H/\left( 6m_{e}cE_{F}\right) $) and its linear increase as a
function of the magnetic field in weak enough fields ($h\ll \epsilon $).
However, the saturation of the NE signal at the fields $h\gtrsim \epsilon $
predicted by the GL model does not find its confirmation in the experimental
data. Quite contrarily, the experiments show that both conventional and
unconventional superconductors demonstrate the characteristic maximum of the
NE signal at $h\sim \epsilon $ \cite{T12,L14}. The maximum in the magnetic
field dependence of the NE signal persists at $T\gg T_{c}$ \cite{T12}, i.e.
far beyond the GL model range of validity.

It worth to recall that the similar problem was discussed in 1970's in
relation to the formal saturation of the fluctuation magnetization of 2D
superconductors in strong fields calculated using the GL model \cite%
{GLmagnetis70}. This seeming paradox was explained by the early breakdown of
the GL scenario at relatively weak magnetic fields where the magnetic length
of a Cooper pair approaches the GL coherence length $\xi _{GL}\left(
\epsilon \right) .$ Interestingly, this happens at $h\sim \epsilon $, where
the minimum in magnetization and the maximum in the NE signal magnetic field
dependencies are expected.

Ref. \cite{KAE72} shows that the short wave and dynamic fluctuation modes
must be taken into account when calculating the magnetization and
conductivity of fluctuative superconductors. In full generality, the
fluctuation part of the free energy can be represented as the trace of the
logarithm of the fluctuation propagator \cite{LV09}%
\begin{equation}
\mathcal{F}_{\mathrm{cp}}^{\left( 2D\right) }(T,H)=-\frac{|e|H}{\pi }%
T\sum_{k}\sum_{n}\ln \left[ gL_{n}^{-1}\left( \Omega _{k}\right) \right] ,
\label{freeen}
\end{equation}%
where $g$ is the effective BCS interaction strength, while the fluctuation
propagator $L_{n}\left( \Omega _{k}\right) $ is the two particle Green
function describing fluctuation Cooper pairings of electrons in a wide range
of temperatures above the line $T_{c}\left( H\right) $ \cite{LV09}: 
\begin{equation*}
L_{n}^{-1}\left( \Omega _{k}\right) =-\zeta \mathcal{E}_{n}\left( \Omega
_{k}\right) .
\end{equation*}%
Here $\zeta $ is the electron density of states and 
\begin{equation*}
\mathcal{E}_{n}\left( \Omega _{k}\right) =\ln \frac{T}{T_{c0}}+\psi \left[ 
\frac{1+\left\vert k\right\vert }{2}+\frac{|e|\mathcal{D}H}{\pi cT}\left( n+%
\frac{1}{2}\right) \right] -\psi \left( \frac{1}{2}\right) ,
\end{equation*}
where $\mathcal{D}$ is the electron diffusion coefficient. The summation in
Eq. (\ref{freeen}) is performed over the Landau levels $n$ and corresponding
bosonic frequencies of Cooper pairs $\Omega _{k}=2\pi Tk$.

According to Eq. (\ref{ncpfull}), the magnetization current of fluctuating
Cooper pairs now can be expressed through the second derivative of the free
energy (\ref{freeen}). One can notice that the summation of the terms in Eq.
(\ref{freeen}) containing $\ln \left( g\zeta \right) $ yields temperature
and magnetic field independent constants which do not contribute to the NE
signal. As a result, one can find the general expression for the NE signal
as:

\begin{equation}
N_{\mathrm{cp}}^{\left( \mathrm{2D}\right) }\left( T,H\right) =\frac{c\rho _{%
\mathrm{cp}}^{xx}}{\Phi _{0}}\frac{d}{dT}\frac{\partial }{\partial H}\left[
HT\sum_{k}\sum_{n}\ln \mathcal{E}_{n}\left( \Omega _{k}\right) \right] .
\label{NCPgen}
\end{equation}%
This expression allows finding $N_{\mathrm{cp}}^{\left( \mathrm{2D}\right)
}\left( T,H\right) $ in the whole range of magnetic fields and temperatures
above the phase boundary $H_{c2}\left( T\right) .$

In particular, one can derive analytically the magnetic field and
temperature dependencies of the NE signal at very low temperatures close to
the second critical field $H_{c2}\left( 0\right) $, where only the
contribution of the lowest Landau level is essential and summation over
bosonic frequencies can be done exactly. The corresponding expression for $%
M_{\mathrm{cp}}^{\left( 2D\right) }\left( T,\widetilde{h}\right) $ can be
found in Ref.\cite{GL01}. Of interest for us is its temperature derivative
which has a form: 
\begin{widetext}
\begin{equation}
\frac{dM_{\mathrm{cp}}^{\left( 2D\right) }\left( T,\widetilde{h}\right) }{dT}%
=\frac{|e|}{\pi \gamma _{E}}\left\{ \left( \frac{16\gamma _{E}^{3}T^{2}}{\pi
^{2}\widetilde{h}T_{c0}^{2}}-1\right) \left[ \frac{\gamma _{E}}{\widetilde{h}%
}-\frac{\widetilde{h}T_{c0}^{2}}{2\gamma _{E}T^{2}}\psi ^{\prime }\left( 
\frac{\widetilde{h}T_{c0}}{2\gamma _{E}T}\right) \right] -\frac{T_{c0}}{T}%
\right\} .  \label{dmdT}
\end{equation}%
\end{widetext}$\gamma _{E}=1.78.$ Here we took into account also the
temperature dependence of the second critical field: 
\begin{equation*}
H_{c2}\left( T\right) =\frac{\pi T_{c0}}{2\gamma _{E}\mathcal{D}}\frac{c}{|e|%
}\left( 1-\gamma _{E}^{2}\frac{T^{2}}{T_{c0}^{2}}\right) .
\end{equation*}%
The Cooper pair contribution to the resistivity above $T_{c}$ can be found
in Ref. \cite{GL01,GVV11}: 
\begin{equation}
\rho _{\mathrm{cp}}^{\left( 2D\right) }\left( T,\widetilde{h}\ll 1\right)
=\left\{ 
\begin{array}{c}
\frac{\pi ^{2}}{2\gamma _{E}e^{2}}\frac{\widetilde{h}T_{c0}}{T},\qquad 
\widetilde{h}\ll \frac{T}{T_{c0}} \\ 
\frac{3\pi ^{2}}{2e^{2}\ln \widetilde{h}},\qquad \frac{T}{T_{c0}}\ll 
\widetilde{h}%
\end{array}%
\right\vert .  \label{cond}
\end{equation}%
with $\widetilde{h}=\left[ H-H_{c2}\left( 0\right) \right] /H_{c2}\left(
0\right) .$ Eqs. (\ref{dmdT}) and (\ref{cond}) allow us to obtain the
asymptotic behavior of the Nernst signal in the low temperature range: 
\begin{equation*}
N_{\mathrm{cp}}^{\left( 2D\right) }\left( T,\widetilde{h}\right) =\frac{%
\gamma _{E}}{|e|}\left( \frac{T}{\widetilde{h}T_{c0}}\right) \left\{ 
\begin{array}{c}
\frac{8\gamma _{E}}{\pi },\qquad \widetilde{h}\ll \frac{T^{2}}{T_{c0}^{2}}
\\ 
-\frac{1}{\widetilde{h}\ln \frac{1}{\widetilde{h}}},\qquad \frac{T}{T_{c0}}%
\ll \widetilde{h}%
\end{array}%
\right\vert .
\end{equation*}

The temperature dependence of the maximum\ in the NE signal dependence on
the magnetic field is of special interest. Recently, the authors of Ref. 
\cite{T12,L14} have proposed using it for the precise determination of the
second critical field $H_{c2}\left( 0\right) $, often unaccessible for
direct measurements because of its huge value. The analysis of the
experimental data obtained on the HTS compound $Pr_{2-x}Ce_{x}CuO_{4}$ led
the authors of \ Ref. \cite{T12} to propose a phenomenological expression:

\begin{equation}
H_{\max }^{\ast }(T)=H_{c2}\left( 0\right) \ln \frac{T}{T_{c0}}.  \label{T}
\end{equation}%
Our Eq. (\ref{NCPgen}) unfortunately does not allow to extract analytically
the temperature dependence of interest, $H_{\max }^{\ast }(T).$
Nevertheless, due to the specific scaling form of Eq. (\ref{NCPgen}) \ the
temperature dependence of the magnetic field corresponding to the maximum of
the Nernst signal can be expressed in the generic form: 
\begin{equation}
H_{\max }^{\ast }\left( \frac{T}{T_{c0}}\right) =\frac{T}{T_{c0}}\varsigma
\left( \ln \frac{T}{T_{c0}}\right) ,  \label{ht*}
\end{equation}%
where $\varsigma \left( x\right) $ is some smooth function which satisfies
the condition $\varsigma \left( 0\right) =0.$

We note that Eq. (\ref{ht*}) coincides with Eq. (\ref{T}) \ only in the
particular case of $\varsigma \left( x\right) =x\exp \left( -x\right) .$ In
the case of any other analytical function $\varsigma \left( x\right) $, the
magnetic field corresponding to the maximum of the NE signal, $H_{\max
}^{\ast }(T),$ would increase linearly with the increase of temperature. The
heuristic justification of Eq. (\ref{T}) \ is based on the statement that
the maximum in the NE signal magnetic field dependence occurs where the FCP
size $\xi _{GL}(T)$ is of the order of its magnetic length $\ell _{H_{\max
}^{\ast }}=\left( c/|e|H_{\max }^{\ast }\right) ^{1/2}$ \cite{Aub06,T12,L14}%
. Close to the critical temperature, this indeed yields $H_{\max }^{\ast
}\sim H_{c2}\left( 0\right) \left( T-T_{c0}\right) /T_{c0}.$ Far from $%
T_{c0},$ the authors of \cite{Aub06,T12,L14} extend the GL expression as $%
\xi _{GL}(T)=\xi _{BCS}/\sqrt{\ln \frac{T}{T_{c0}}\text{ }},$which brings
them to Eq. (\ref{T}). We believe that this extension lacks justification,
and the rigorous expression (\ref{NCPgen}) needs to be used, in the general
case.

In conclusion, we have derived the Seebeck and Nernst-Ettingshausen
coefficients for fluctuative superconductors, in the local equilibrium
approximation. A thermodynamical approach allows analytical evaluation of
both constants which appear to be different by orders of magnitude due to
the crucial contribution of magnetization currents to the NE signal. We
explain the non-monotonous behaviour of the NE signal as a function of
magnetic field above $T_{c}$ and estimate the position of the maximum of
this dependence from a simple scaling argument.

\textit{Acknowledgement.} We thank Yuriy Galperin for the critical
reading of the manuscript and useful comments and 
Andreas Glatz for fruitful discussions,

\end{document}